\documentclass[reprint,amsmath,amssymb,aps,prb]{revtex4-2} 
\usepackage{graphicx, color}
\usepackage{dcolumn}
\usepackage{bm}
\usepackage{ulem}
\usepackage[colorlinks,urlcolor=blue,linkcolor=blue,anchorcolor=blue,citecolor=blue]{hyperref} 

\begin{document}
\preprint{APS/123-QED}
\title{
Bipolarized Weyl semimetals and quantum crystal valley Hall effect in two-dimensional altermagnetic materials}
\author{Chao-Yang Tan$^{1}$}
\author{Ze-Feng Gao$^{1}$}
\author{Huan-Cheng Yang$^{1}$}
\author{Kai Liu$^{1}$}
\author{Peng-Jie Guo$^{1}$}
\email{guopengjie@ruc.edu.cn}
\author{Zhong-Yi Lu$^{1,2}$}
\email{zlu@ruc.edu.cn}
\affiliation{1. School of Physics and Key Laboratory of Quantum State Construction and Manipulation (Ministry of Education). Renmin University of China, Beijing 100872, China}
\affiliation{2. Hefei National Laboratory, Hefei 230088, China}

\begin{abstract}

Magnetism and topology are two major areas of condensed matter physics. The combination of magnetism and topology gives rise to more novel physical effects, which have attracted strongly theoretical and experimental attention. Recently, the concept of altermagnetism has been introduced, characterized by a dual nature: real-space antiparallel spins with zero total magnetic moment and reciprocal-space anisotropic spin polarization. The amalgamation of altermagnetism with topology may lead to the emergence of previously unobserved topological phases and the associated physical effects. In this study, utilizing a four-band lattice model that incorporates altermagnetism and spin group symmetry, we demonstrate that type-I, type-II, and type-III bipolarized Weyl semimetals can exist in altermagnetic systems. Through the first-principles electronic structure calculations, we predict four ideal two-dimensional type-I altermagnetic bipolarized Weyl semimetals Fe$_2$WTe$_4$ and Fe$_2$MoZ$_4$ ({Z=S, Se, Te}). More significantly, we introduce the quantum crystal valley Hall effect, a phenomenon achievable in three of these materials namely Fe$_2$WTe$_4$, Fe$_2$MoS$_4$, and Fe$_2$MoSe$_4$, when spin-orbit coupling is considered. Therefore, our work not only enriches the topological phases but also provides a material platform for studying novel topological phases in altermagnetism.

\end{abstract}

\maketitle

\section{Introduction}
\begin{figure}[htbp]
	\centering
	\includegraphics[width=8.6cm]{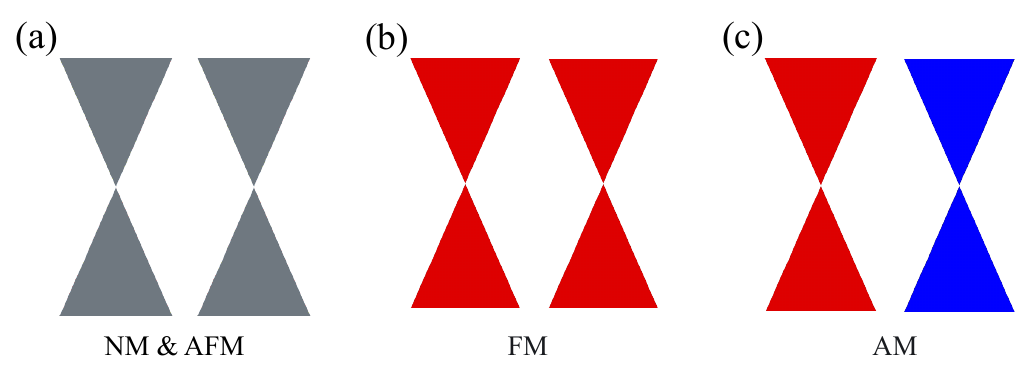}
	\caption{The schematic diagrams of three different Weyl semimetal phases. (a) Nonmagnetic (NM) or antiferromagnetic (AFM) Weyl semimetal; (b) Ferromagnetic (FM) semimetal; (c) bipolarized altermagnetic (AM) semimetal. The gray cones represent Weyl points without spin polarization. The red and blue cones represent spin-up and spin-down polarized Weyl points, respectively.}
	\label{fig1}
\end{figure}

The topological Weyl semimetals have attracted much attention due to their novel physical properties, such as topological protected Fermi arc, chiral zero sound, chiral anomaly, large magnetoresistance effect, large intrinsic anomalous Hall effect \cite{WSMs-1,WSMs-2,WSMs-3,WSMs-4,WSMs-5,WSMs-6,WSMs-7,WSMs-8,WSMs-9,WSMs-10,TWNR-2021,TVHE-2022,WPtype2-2024,TypeIINat2015,TypeIIIPRB2021,TypeIIIPRB2024,TypeIIIPRB2025}, etc. Since Weyl points are formed by linear crossing of  two nondegenerate energy bands, Weyl semimetals can only exist in the materials without the joint symmetry of time reversal and space inversion. The Weyl points of nonmagnetic or conventional antiferromagnetic Weyl semimetals have no spin polarization, especially in the Weyl semimetals without spin-orbit coupling (SOC) \cite{AFMWS-1,AFMWS-2,AFMWS-3,AFMWS-4}  (Fig.~\ref{fig1}{(a)}), while the Weyl points of ferromagnetic Weyl semimetals have only spin-up or spin-down polarization (Fig.~\ref{fig1}{(b)}). This then raises an intriguing question: Can a Weyl semimetal accommodate both spin-up and spin-down polarized Weyl points simultaneously? (Fig.~\ref{fig1}{(c)}).

Recently, based on spin group theory, a new magnetic phase, namely altermagnetism defined in the non-relativistic case, which is different from ferromagnetism and conventional antiferromagnetism, has been proposed \cite{altermagnetism-1,altermagnetism-2,altermagnetism-3,altermagnetism-4,altermagnetism-PRX-1,altermagnetism-PRX-2,altermagnetism-PRX-3}. Due to the duality of real-space antiparallel spins with zero total magnetic moment and reciprocal-space anisotropic spin polarization, altermagnetic (AM) materials show many novel physical effects, including spin-splitting torque \cite{SST-PRL2021,SST-NE2022,SST-PRL2022,SST-PRL2022-2}, giant magnetoresistance effect \cite{GMR-PRX2022}, tunneling magnetoresistance effect \cite{GMR-PRX2022,TMR-Shao2021}, piezomagnetic effect \cite{piezomagnetism-NC}, nontrivial superconductivity \cite{SC-AM, MCM-liu2023}, time-reversal odd anomalous effect \cite{AHE-Sinova2022,AHE-RuO2-NE2022,AHE-MnTe-PRL2023,AHE-hou2023,altermagnetism-2,MOE-Yao2021,CTHE-Yao2024,2Dmodel-arxiv-1,2Dmodel-arxiv-2}, quantum anomalous Hall effect \cite{QAH-npj2023}, higher-order topological states \cite{HighoT-liu2024}, altermagnetic ferroelectricity \cite{LiFe2F6-guo2023}, strong spin-orbit coupling effect in light element altermagnetic materials \cite{NiF3-qu2024} and so on. Moreover, the already predicted altermagnetic materials cover metals, semimetals, and insulators, which provides a guarantee for the realization of many novel physical effects in experiments \cite{AI-gao2023,SSG-song2024,YJZ-2024,liu-arxiv2024,AM-nature2024,AM-MnTe2024,AM-CrSb2024,RuO2-2024}. More importantly, the reciprocal-space anisotropic spin polarization of altermagnetism also provides possibility to realize a topological Weyl semimetal with both spin-up and spin-down polarized Weyl points at the same energy(Fig.~\ref{fig1}{(c)}). Here, we term this class of Weyl semimetals as bipolarized Weyl semimetals (BWSs).

Moreover, Weyl-like fermions are classified by emergent Lorentz symmetry and the topology of their Fermi surfaces. Type-I Weyl semimetals preserve Lorentz symmetry and possess point-like Fermi surfaces, whereas type-II Weyl semimetals host overtilted Weyl cones with touching electron and hole pockets, thereby breaking the symmetry \cite{WSMs-1,TypeIINat2015}. At the critical transition between them lies the type-III Weyl semimetal, featuring a critically tilted Weyl cone and a line-like Fermi surface \cite{TypeIIIPRB2021,TypeIIIPRB2024,TypeIIIPRB2025}. This classification can be naturally generalized to BWSs, suggesting the possible realization of type-I, type-II, and type-III BWSs in altermagnetic systems.

In this work, based on spin group symmetry analysis, lattice model, and the first-principles electronic structure calculations, we not only demonstrate that type-I, type-II, and type-III BWS can be realized in AM systems, but also predict four ideal two-dimensional type-I BWSs Fe$_2$WTe$_4$ and Fe$_2$MoZ$_4$ ({Z=S,Se,Te}). Furthermore, we find quantum crystal valley Hall effect in Fe$_2$WTe$_4$, Fe$_2$MoS$_4$, and Fe$_2$MoSe$_4$ under SOC. And they can transform from quantum crystal valley Hall phase to Chern insulator phase under strain. In contrast, Fe$_2$MoTe$_4$ remains to be a Weyl semimetal with only one pair of Weyl points under SOC.

\section{ Lattice Model} 
\begin{figure}[htbp]
	\centering
	\includegraphics[width=8.6cm]{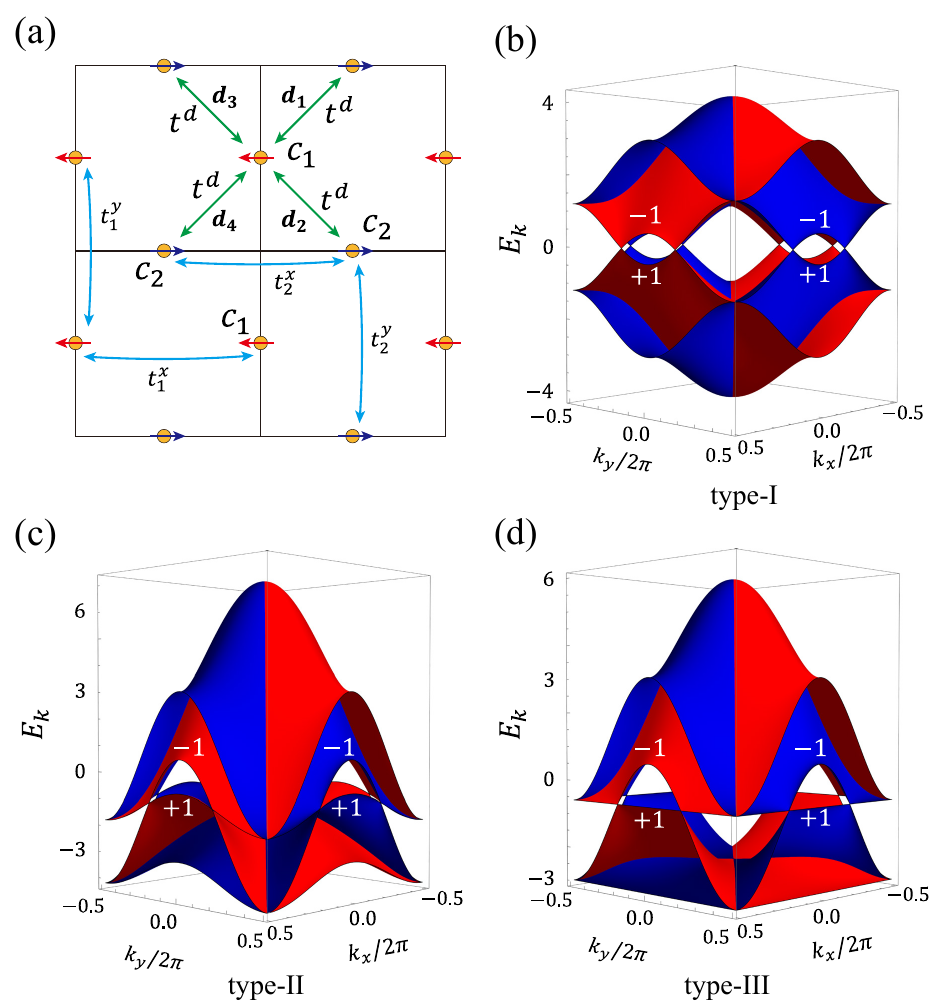}
	\caption{ Two-dimensional lattice model with altermagnetism. (a) The schematic illustration for the hoppings between lattice sites for Hamiltonian Eq. (\ref{Eq1}). The red and blue arrows represent spin-up and spin-down magentic moments, respectively. (b), type-I BWS with $t_1=-t_2=0.4$, (c), type-II BWS with $t_1=0.9$, $t_2=0.0$, (d), type-III BWS with $t_1=1.2$, $t_2=0.3$. Other parameters: $t^d=1$, $|\textbf{m}|=m_z = 1.2$. The red and blue represent spin-up and spin-down bands, respectively. The 1 and $-1$ represent eigenvalues of spin symmetry $\left\{ E||C_{2x} \right\}$ or $\left\{ E||C_{2y} \right\}$. }
	\label{fig2}
\end{figure}

We construct a square lattice containing two sites (labeled by sublattice index $\alpha=1,2$) within a unit cell (Fig.~\ref{fig2}(a)). 
The corresponding tight-binding (TB) model without SOC is

\begin{align}
H = & \sum_{d_i,j} \left[ t^{d} C_{1,j}^\dagger C_{2,j+\bm{d}_i} + \mathrm{H.c.}\right] \notag\\
    &+\sum_{\alpha,j} \left[ t^{x}_{\alpha} C_{\alpha,j}^\dagger C_{\alpha,j+\mathbf{x}} + t^{y}_{\alpha} C_{\alpha,j}^\dagger C_{\alpha,j+\mathbf{y}} + \mathrm{H.c.}\right] \notag\\
    &+\sum_{\alpha,j} C_{\alpha,j}^\dagger \left[\bf{m}_\alpha \cdot \boldsymbol{\sigma} \right] C_{\alpha,j}
    \label{Eq1}
\end{align}
where $ C_{\alpha,j}^\dagger=\left( C_{\alpha,j\uparrow}^\dagger, C_{\alpha,j\downarrow}^\dagger \right) $ and $ C_{\alpha,j}=\left( C_{\alpha,j\uparrow}, C_{\alpha,j\downarrow} \right) $ represent electron creation and annihilation operators, respectively. The terms containing $t^{d}$ and $t^{x,y}_\alpha$ represent the nearest and next nearest hopping, respectively. As illustrated in Fig.~\ref{fig2}(a), $\bm{d}_{1,4}=\pm\frac{1}{2}(\mathbf{x}+\mathbf{y}), \bm{d}_{2,3}=\pm\frac{1}{2}(\mathbf{x}-\mathbf{y})$ and $\mathbf{x}=a_1 \hat{x}, \mathbf{y}=a_2 \hat{y}$ denote the direction of hopping, where $a_1, a_2$ are the lattice constants along the unit vectors $\hat{x}, \hat{y}$. The $\sigma_0$ and $\boldsymbol{\sigma}$ are identity matrix and Pauli matrix, respectively. The $\textbf{m}_\alpha=(-1)^\alpha\textbf{m}$ stands for the static collinear AFM order, which couples with the electron spin as a Zeeman field. Due to the two opposite spin sites located at the space-inversion invariant positions, the collinear AFM order breaks spin symmetry $\left\{C^{\bot}_2||I\right\}$ [Sec. I of the Supplemental Material (SM)]. Moreover, since $t_1^{x,y}\neq t_2^{x,y}$ breaks the spin symmetry $\left\{C^{\bot}_2||\tau\right\}$ but $t_1^{x,y} = t_2^{y,x}$ preserves spin symmetry $\left\{C^{\bot}_2||C_{4z}\right\}$. Thus, this lattice model is a \textit{d}-wave altermagnetism.

After performing the Fourier transformation, the TB Hamiltonian can be rewritten as $H=\sum_k\psi_k^\dagger H_0 \psi_k$ in momentum space with basis $\psi^\dagger = \left( C_{1k\uparrow}^\dagger, C_{1k\downarrow}^\dagger, C_{2k\uparrow}^\dagger, C_{2k\downarrow}^\dagger \right)$, where $H_0$ reads
\begin{align}
	H_0 =\Gamma^{+}_k \tau_0 \sigma_0 + \Gamma_k^{12} \tau_x \sigma_0+\Gamma_k^{-} \tau_z \sigma_0 - \tau_z \bf{m} \cdotp \boldsymbol{\sigma}
 \label{Eq2}
\end{align}
in which Pauli matrix $\boldsymbol{\tau}$ is the pseudospin matrix in lattice space. The auxiliary functions $\Gamma_k^{12}=4t^d \cos\frac{k_x}{2} \cos\frac{k_y}{2}$ and $\Gamma_k^{\pm}=(t_1 \pm t_2)\cos k_x + (t_2 \pm t_1)\cos k_y$ with $t_\alpha^x=t_\alpha$  are related to inter and intra sublattice hoppings, respectively. Altermagnetism leads to anisotropic spin splitting, so what does the size of spin splitting depend on? The analysis of the lattice model shows the size of the spin splitting is determined by the strength of anisotropic hopping $|t_1-t_2|$ (Detailed calculations and analyzes are presented in the SM). 

When $2|t_1-t_2|>|\mathbf{m}|$, the band crossing occurs, which is a necessary condition for the realization of BWSs. Further symmetry analysis shows that there are two crossing points with spin-up polarization and two other crossing points with spin-down polarization protected by spin symmetry $\left\{ E||C_{2y} \right\}$ and $\left\{ E||C_{2x} \right\}$, respectively. Therefore, a BWS can be realized in an AM system. Interestingly, by adjusting the hopping terms $t_1$ and $t_2$, we are able to achieve type-I, type-II, and type-III BWSs which are shown in Fig.~\ref{fig2}{(b), (c), and (d)}, respectively.

\section{Candidate materials} 
\begin{figure}[htbp]
	\centering
	\includegraphics[width=8.6cm]{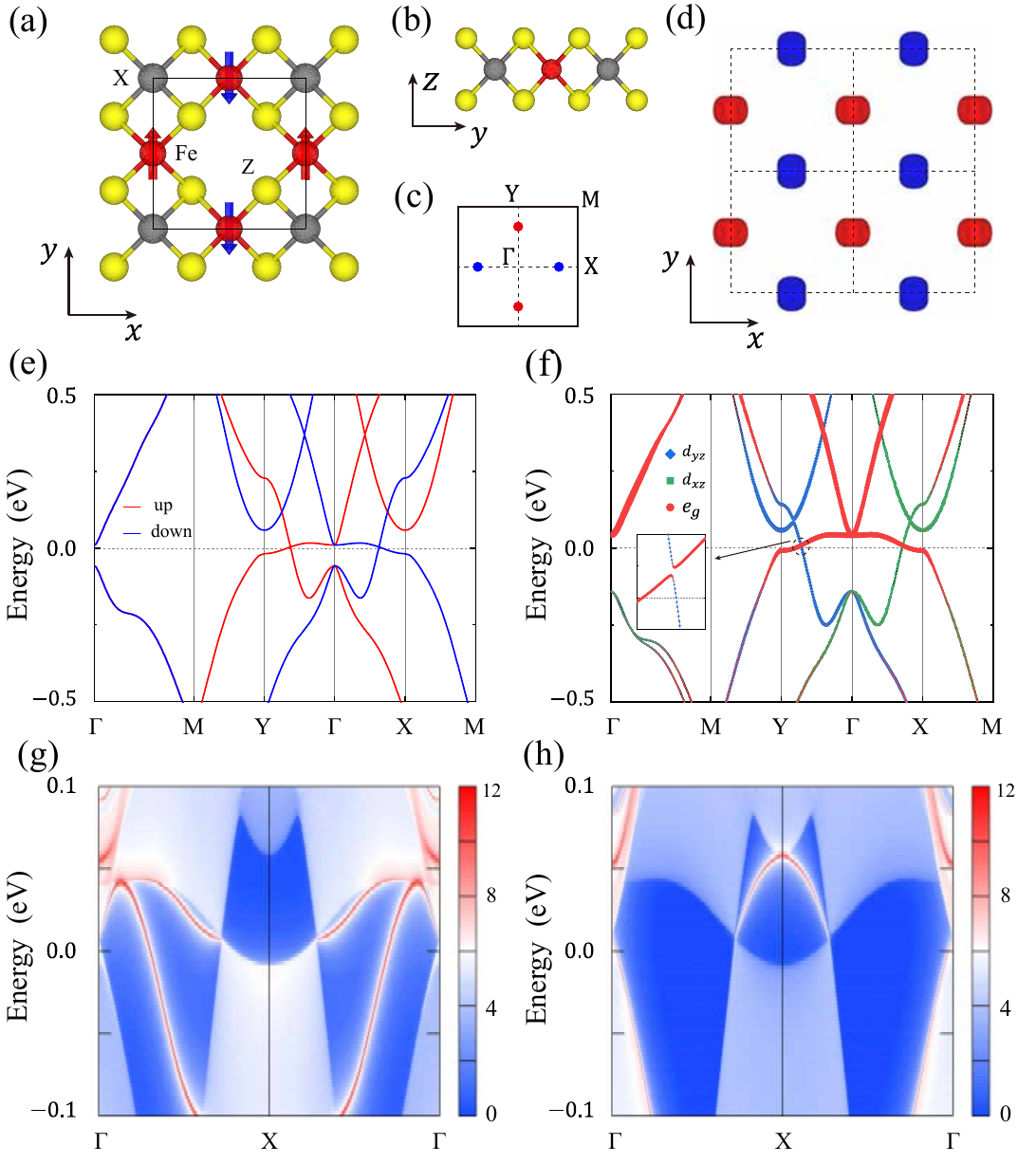}
	\caption{Crystal and magnetic structures of monolayer $\rm{Fe_2XZ_4}$ and electronic properties of monolayer $\rm{Fe_2WTe_4}$. (a) and (b) represent the top and side views of crystal and magnetic structure of $\rm{Fe_2XZ_4}$, respectively. The red and blue arrows represent spin-up and spin-down magnetic moments, respectively. (c) The BZ of $\rm{Fe_2XZ_4}$ where the high-symmetry points are labeled. The red and blue points represent spin-up and spin-down polarized Weyl points, respectively. (d) The polarization charge density of $\rm{Fe_2WTe_4}$ without SOC (red: spin-up, blue:spin-down).The electronic structure along the high-symmetry directions without SOC (e) and with SOC (f). The $e_g$ represents the $d_{z^2}$ and $d_{x^2-y^2}$ orbitals of Fe element. The Spectral function of monolayer $\rm{Fe_2WTe_4}$ for a semi-infinite termination with Fe-W atoms left boundary (g) and Te atoms right boundary (h). }
	\label{fig3}
\end{figure}

Monolayer $\rm{Fe_2XZ_4}$ (Fe$_2$WTe$_4$ and Fe$_2$MoZ$_4$) takes a square lattice structure with the symmorphic space group $P$-$42m$ (No.111) symmetry and the corresponding point group is $D_{2d}$ with generators $S_{4z}$ and $C_{2x}$.  
Monolayer $\rm{Fe_2XZ_4}$ contains three atomic layers where Fe$_{2}$X atomic layer is sandwiched by two Z atomic layers as shown in Fig.~\ref{fig3}{(a) and (b)}, and the corresponding BZ is shown in Fig.~\ref{fig3}{(c)}. Moreover, the dynamical stability of monolayer $\rm{Fe_2XZ_4}$ is confirmed by the phonon calculations (Fig.~S8) \cite{SM-1}. Since the crystal structure of monolayer $\rm{Fe_2XZ_4}$ is proposed based on the already synthesized layered $\rm{Cu_2XZ_4}$  (Z=S,Se) \cite{Cu2MX4-1993,Cu2MX4-2005,CuMoS-2012,Cu2MX4-2019,Cu2MX4-2022} and $\rm{Ag_2WS_4}$ \cite{Ag2WS4-2018}, monolayer $\rm{Fe_2XZ_4}$ may be synthesized experimentally.

To determine the magnetic ground states of monolayer $\rm{Fe_2XZ_4}$, we consider three likely magnetic structures including ferromagnetic, altermagnetic (AFM1)and zigzag antiferromagnetic (AFM2) (Fig.~S9). The calculated results show that the AM ordering (Fig.~\ref{fig3}{(a)}) is always magnetic ground state for all four materials $\rm{Fe_2XZ_4}$, as shown in Table.~S2. From Fig.~\ref{fig3}{(a)}, the magnetic and crystal primitive cells are identical, thus $\rm{Fe_2XZ_4}$ has no $\left\{ C_2^{\bot}||\tau \right\}$ spin symmetry. And due to the lack of space-inversion symmetry, $\rm{Fe_2XZ_4}$ must not have $\left\{ C_2^{\bot}||I \right\}$ spin symmetry. Meanwhile, the $\rm{Fe_2XZ_4}$ has $\left\{ C_2^{\bot}||S_{4z} \right\}$ spin symmetry. Thus, all the $\rm{Fe_2XZ_4}$ are \textit{d}-wave AM materials. In the following, we present nontrivial topological states with $\rm{Fe_2WTe_4}$ as a representative material, while the calculated results of other three materials are presented in the SM.

For monolayer $\rm{Fe_2WTe_4}$, the W atom is at the corner of primitive cell and there is no W atom in the center, which leads to a strongly anisotropic polarized charge density around the magnetic Fe atom as shown in Fig.~\ref{fig3}(d).  The strongly anisotropic polarized charge density results in significantly different hopping interactions in x or y direction between the next nearest Fe atoms, which can lead to the large spin splitting in the nonrelativistic case according to our lattice model analysis.
Just like our analysis, the spin splitting caused by the exchange interactions in altermagnetic $\rm{Fe_2WTe_4}$ is very large. The spins of these bands along the $\Gamma$-X and $\Gamma$-Y are opposite, reflecting the characteristics of $d$-wave altermagnetism (Fig.~\ref{fig3}{(e)}). More significantly, altermagnetic $\rm{Fe_2WTe_4}$ is a perfect semimetal with only four doubly degenerate crossing points at the Fermi level (Fig.~\ref{fig3}{(c) and (e)}). These crossing points can be considered as a valley degree of freedom, similar to the case of graphene. Manipulating this valley degree of freedom will give rise to novel effects as shown below. The further orbital weight analysis shows that the two crossing bands along the $\Gamma$-Y direction are respectively contributed by the $e_{g}$ and $d_{yz}$ orbitals of Fe atoms (Fig.~\ref{fig3}{(f)}). Since the $e_{g}$ orbital is invariable but the $d_{yz}$ orbital changes to $-d_{yz}$ under the $\left\{ E||C_{2y} \right\}$ operation, the corresponding eigenvalues of the two crossing bands are 1 and $-1$, respectively. Thus, the two crossing points on the $\Gamma$-Y direction are the Weyl points protected by $\left\{ E||C_{2y} \right\}$ spin symmetry. Likewise, the two crossing points in the $\Gamma$-X direction are also the Weyl points protected by $\left\{ E||C_{2x} \right\}$ spin symmetry. Interestingly, the two Weyl points on the $\Gamma$-Y direction have spin-up polarization, while the two Weyl points in the $\Gamma$-X direction have spin-down polarization (Fig.~\ref{fig3}{(e)}). Thus, the monolayer AM $\rm{Fe_2WTe_4}$ is a bipolarized Weyl
semimetal.

With SOC, monolayer $\rm{Fe_2WTe_4}$ changes from spin group symmetry to magnetic group symmetry. According to our magnetic anisotropy calculations, the direction of the Néel vector is along the $z$-axis, which has magnetic point group symmetry 2$S_{4z}T$, $C_{2z}$, $C_{2x}T$, $C_{2y}T$, and 2$\sigma_d$. The symmetry of $\Gamma$-Y axis changes from the $\left\{ E||C_{2y} \right\}$ spin symmetry to the $C_{2x}T$ symmetry. Both $e_g$ and $d_{yz}$ orbitals are invariant under the  $C_{2x}$ operation, so the two crossing bands in the $\Gamma$-Y direction have the same irreducible representation, thus opening a gap (2.5 meV) (inset of Fig.~\ref{fig3}{(f)}). Likewise, the two Weyl points in the $\Gamma$-X direction also open a gap of 2.5 meV. Since the SOC around the Fermi level for the Fe$_2$WTe$_4$ family is very weak, the Fe$_2$WTe$_4$ family is good candidates for BWSs. 

Due to the $C_{2z}$ symmetry, the two Weyl points in the $\Gamma$-X or $\Gamma$-Y direction contribute the same Berry curvature, but the two pairs of Weyl points in the $\Gamma$-X and $\Gamma$-Y directions contribute the opposite Berry curvature due to the $S_{4z}T$ symmetry (Fig.~\ref{fig4}{(a)}). We know that a two-dimensional linear masssive Weyl points contributes $\pi$ Berry phase, so a pair of masssive Weyl points on the $\Gamma$-X ($\Gamma$-Y) direction contributes -2$\pi$ (2$\pi$) Berry phase, which corresponds to the Chern number -1 (1). Furthermore, the opposite Chern number may lead to nontrivial edge states with opposite chirality in a boundary, which have been confirmed by the calculations (Fig.~\ref{fig3}{(g) and (h)}). Since the SOC of $\rm{Fe_2WTe_4}$ is very weak, these two opposite chiral edge states are not easily distinguishable in Fig.~\ref{fig3}{(g) and (h)}. To better illustrate it, we also construct a lattice model with SOC to calculate the chiral edge states, which are shown in Fig.~S4(b). On the other hand, if the band inversion along the $\Gamma$-X direction disappears by applying tensile stress, leaving only a pair of Weyl points along the $\Gamma$-Y direction, this will result in a Chern insulator phase with Chern number 1 under SOC. If the band inversion along the $\Gamma$-Y direction disappears by applying tensile stress, leaving only a pair of Weyl points along the $\Gamma$-X direction, this will result in a Chern insulator phase with Chern number -1 (Fig.~S17). Therefore, applying tensile stress in different directions can realize different Chern insulator phases in $\rm{Fe_2WTe_4}$.

\section{Quantum crystal valley Hall effect}
\begin{figure}[htbp]
\centering
\includegraphics[width=8.6cm]{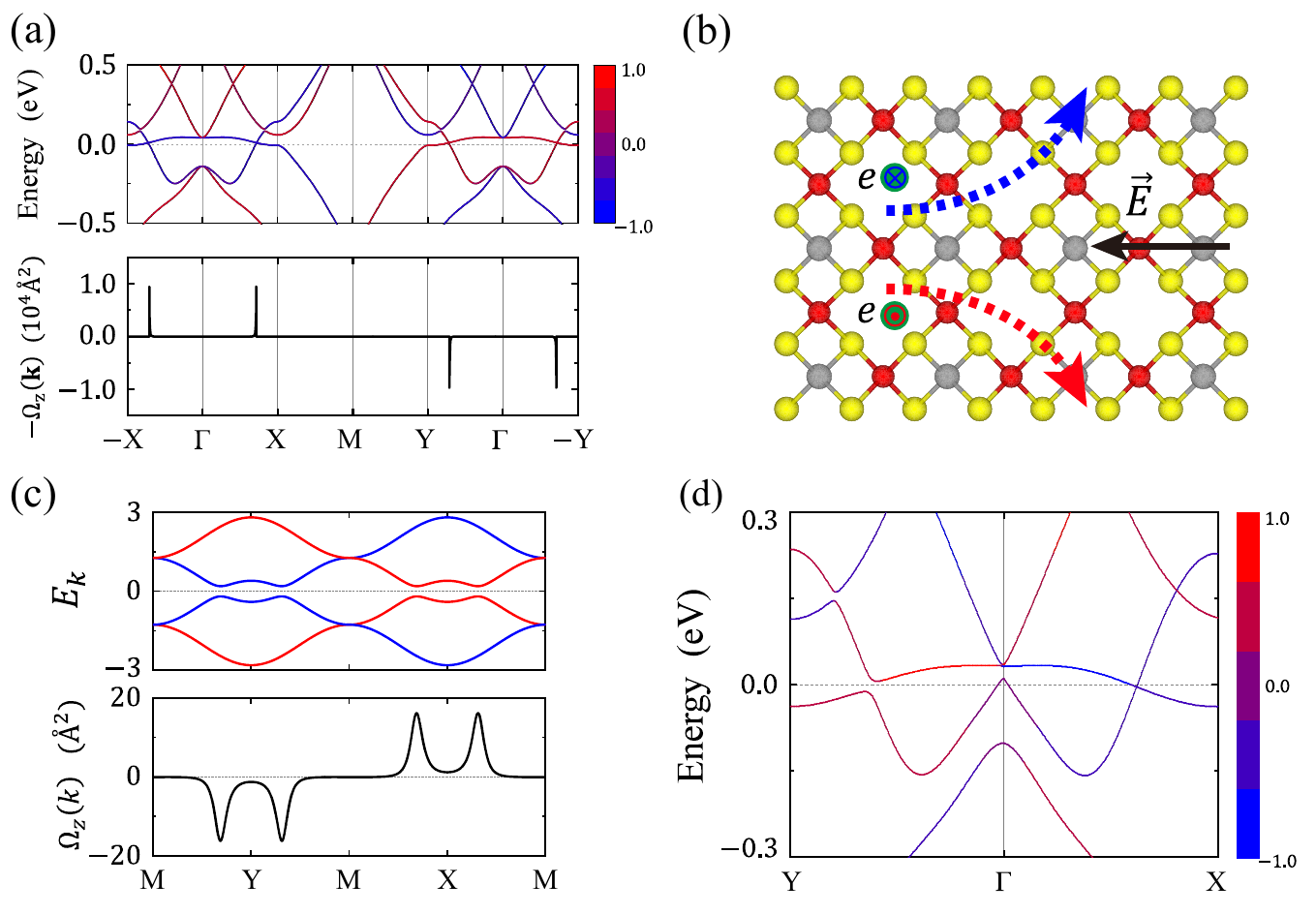}
\caption{(a) Top panel: the electronic band structure of monolayer $\rm{Fe_2WTe_4}$ with the spin projection $s_z$; bottom panel: the corresponding berry curvature along high-symmetry line. (b) The schematic diagram of topological valley Hall effect in monolayer $\rm{Fe_2WTe_4}$. The $\vec{E}$ and  $e$ represent external electric field and electrons, respectively. The red and blue arrows (out of plane) represent up and down spins, respectively. (c) Top panel: the band structure of TB model with SOC, bottom panel: the corresponding berry curvature along high-symmetry line. (d) the electronic structure for Fe$_2$MoTe$_4$ with SOC.} 
\label{fig4}
\end{figure}

For two-dimensional nonmagnetic materials, the time-reversal symmetry results in energy degeneracy, opposite pseudospin and opposite Berry curvature respectively at momenta $k$ and $–k$ with SOC, that is $E_\uparrow (k)=E_\downarrow (-k)$ and $\Omega_z (k)=-\Omega_z (-k)$. In the presence of an in-plane electric field, there is an anomalous velocity $\boldsymbol{v}_a \sim \boldsymbol{E}\times \boldsymbol{\Omega} (k)$ due to Berry curvature \cite{BerryPhase-PRM2010}. Since both the pseudospins and the Berry curvature at valley $K$ and $-K$ are opposite, valley electrons at $K$ and $-K$ move in the opposite direction perpendicular to the electric field, which is the valley Hall effect in two-dimensional nonmagnetic materials, such as MoS$_2$ \cite{MoS2-Science2014,VHE-MoS2-2012}. On the other hand, in two-dimensional conventional antiferromagnetic (CAFM) materials, equivalent time-reversal symmetry exists, namely $IT$ and $T\tau$ symmetries. Although CAFM materials break $I$ and $T$ symmetries, they can possess $IT$ or $T\tau$ symmetry. Since $IT$ symmetry makes the Berry curvature zero at each $k$ point, the valley Hall effect cannot be realized in two-dimensional CAFM materials with $IT$ symmetry. However, $T\tau$ symmetry can also lead to $E_\uparrow (k)=E_\downarrow (-k)$ and $\Omega_z (k)=-\Omega_z (-k)$, so the valley Hall effect can be realized in two-dimensional CAFM materials with $T\tau$ symmetry. Therefore, $T$ or $T\tau$ symmetry is considered a necessary condition for realizing the valley Hall effect. 
Different from two-dimensional nonmagnetic or CAFM materials, two-dimensional altermagnetic materials have no $T$ or $T\tau$ symmetry, but have $M_{xy}$ or $S_{4z}T$ symmetry when considering SOC. Both $M_{xy}$ and $S_{4z}T$ symmetry can lead to energy degeneracy, opposite pseudospin and opposite Berry curvature at high-symmetry $X$ and $Y$ points, that is $E_\uparrow (X)=E_\downarrow (Y)$ and $\Omega_z (X)=-\Omega_z (Y)$ (Fig.~\ref{fig4}{(a)}), which is similar to $T$ and $T\tau$ symmetries. Therefore, the valley Hall effect can also be realized in AM materials with broken $T$ and $T\tau$ symmetries. Since crystal symmetry is essential, we refer to this type of valley Hall effect as the crystal valley Hall effect, which is a novel phenomenon (Fig.~\ref{fig4}{(b)}). 

The quantum valley Hall effect was first proposed based on the valley Hall effect in graphene \cite{QVH-Ding, QVHE-Graphene2015}, characterized by Berry phases of 2$\pi$ and -2$\pi$ at the two valleys ($K$ and $-K$), corresponding to Chern numbers C$_K$ = 1 and C$_{-K}$  = -1. Similar to graphene, in the altermagnetic Fe$_2$WTe$_4$, the valleys polarized with spin up and those with spin down contribute Chern numbers 1 and -1, respectively. Meanwhile, we have also calculated the band structure of the TB model with SOC, which is shown in Fig.~\ref{fig4}{(c)}. From Fig.~\ref{fig4}{(c)}, the valleys with opposite polarizations contribute opposite Chern numbers 1 and -1 originating from the characteristics of the Weyl points with gap. Therefore, the quantum crystal valley Hall effect can be realized both in AM systems. Furthermore, Table S6 in the SM summarizes the symmetry-allowed layer groups that support the quantum crystal valley Hall effect in two-dimensional square lattices, along with their corresponding key symmetry operations.

On the other hand, our calculations show that the other three two-dimensional altermagnetic materials are also BWSs without SOC. The directions of the Néel vector of both $\rm{Fe_2MoS_4}$ and $\rm{Fe_2MoSe_4}$  are the same as that of $\rm{Fe_2WTe_4}$ under SOC, so the quantum crystal valley Hall effect can also be realized in both $\rm{Fe_2MoS_4}$ and $\rm{Fe_2MoSe_4}$ (Figs.~S14 and S15). Meanwhile, both $\rm{Fe_2MoS_4}$ and $\rm{Fe_2MoSe_4}$ also transform from quantum crystal valley Hall insulator phase to Chern insulator phase under strain (Fig.~S17). 

Different from the case of $\rm{Fe_2MoS_4}$, $\rm{Fe_2MoSe_4}$ and $\rm{Fe_2WTe_4}$, the direction of Néel vector of $\rm{Fe_2MoTe_4}$ is along the $x$- or $y$-axis due to the $S_{4}$ symmetry of crystal. When the Néel vector is along the $x$-axis, because there is no any symmetry for the $\Gamma$-Y axis, the pair of spin-up polarized Weyl points in $\rm{Fe_2MoTe_4}$ open gap, while the two spin-down polarized Weyl points in the $\Gamma$-X axis remain stable protected by the $C_{2x}$ symmetry (Fig. \ref{fig4}(d)). That is to say, $\rm{Fe_2MoTe_4}$ is still a Weyl semimetal, but is not a BWS under SOC. If $\rm{Fe_2MoTe_4}$ possesses M$_x$ symmetry, then the spin-up Weyl points along the $\Gamma$-Y direction remain stable even under SOC. In fact, the altermagnetic monolayer CrO exhibits both C$_{2x}$ and M$_x$ symmetries under SOC. Our calculations demonstrate that monolayer CrO is a BWS both in the absence and presence of SOC (Fig. S7). Therefore, BWS can remain stable under SOC.

\section{Conclusion}
In conclusion, we proposed a different topological semimetal phase in altermagnetism, the BWS. There are three main novel phenomena arise from the BWS phase. First, in absence of SOC, the topological edge states of BWS are fully polarized. Specifically, the edge states with spin-up polarization connect the spin-up Weyl points, while those with spin-down polarization connect the spin-down Weyl points. Second, applying uniaxial strain or reorienting the Néel vector breaks specific symmetries, which can regulate the number, polarization, and positions of the Weyl points. Third, when considering SOC, a BWS can transition into an insulator with quantum crystal valley Hall effect.

\section*{Acknowledgments}
This work was financially supported by the National Natural Science Foundation of China (Grant No.12434009, No.12204533, No.62476278 and No.12174443), the National Key R$\&$D Program of China (Grant No. 2024YFA1408601), the Beijing Natural Science Foundation (Grant No.Z200005) and the Innovation Program for Quantum Science and Technology (2021ZD0302402). Computational resources have been provided by the Physical Laboratory of High Performance Computing at Renmin University of China.

\section*{Data availability}
The data that support the findings of this study are available from the corresponding author upon reasonable request.

\nocite{*}

\bibliography{References}

\end{document}